%% file: main.tex
\documentclass[12pt]{iopart}

\usepackage{graphicx}
\usepackage{dcolumn}
\usepackage{bm}
\usepackage{color}

\usepackage{pgfplots}
\usepackage{pgf}
\usepackage{tikz}
\usepackage{graphicx}
 
\begin{document}
\title{Control of magnetism in bilayer CrI$_{3}$ by an external electric field.}
\author{E. Su\'arez Morell }
\ead{eric.suarez@usm.cl}
\author{Andrea Le\'on }
\address{Departamento de F\'isica, Universidad T\'ecnica Federico Santa Mar\'ia, Casilla 110-V, Valpara\'iso, Chile}

\author{R. Hiroki Miwa}
\address{Instituto de F\'isica, Universidade Federal de Uberl\^andia, Caixa Postal 593, 38400-902 Uberlandia, Minas Gerais, Brazil}
\author{P. Vargas}
\address{Departamento de F\'isica, Universidad T\'ecnica Federico Santa Mar\'ia, Casilla 110-V, Valpara\'iso, Chile and \\
Centro para el Desarrollo de la Nanociencia y la Nanotecnolog\'ia,  Santiago, Chile}



\begin{abstract}

Recently intrinsic ferromagnetism in two-dimensional(2D) van der Waals materials was discovered \cite{Huang2017,Gong2017,OHara2018}. A monolayer of Chromiun triiodide(CrI$_3$) is ferromagnetic while a bilayer structure was reported to be anti-ferro magnetic, moreover an external electric field changes its magnetic phase \cite{Jiang2018}. We have studied the two found in nature stackings of CrI$_3$ bilayers and found that indeed the magnetic phase of one of them can be tuned by an external electric field while the other remains ferromagnetic. We simulate those results with \textit{ab initio} calculations and explain them with a simple model based on a rigid shift of the bands associated with different spins. The model can be applied to similar van der Waal stacked insulating bilayer anti-ferromagnets.

\end{abstract}

\maketitle



 Two of the materials are monolayers based on Chromiun atoms, the monolayers were obtained by mechanical exfoliation. One of them is the trihalide CrI$_{3}$ and the other a monolayer of $\rm Cr_{2}Ge_{2}Te_{6}$, their Curie temperature is around 45 Kelvin, too low for most applications but their discovery certainly proved that intrinsic long-range magnetic order can exist in two dimensions. Moreover thin films of $\rm MnSe_{2}$ obtained by molecular beam epitaxy also showed intrinsic ferro-magnetism but at room temperature \cite{OHara2018} and it was also reported strong ferromagnetic order that persist above room temperature for monolayer $\rm VSe_2$, a material that is paramagnetic in the bulk \cite{Bonilla2018}. These finding  are for obvious reasons quite appealing and more proposals of 2D magnetic materials have come out recently \cite{Wu2018,Wang2017}.

One of the reason for the excitement is the relative ease with which two van der Waals materials can be stacked together, that opens a window with endless possibilities in search of wanted properties. In fact magneto optical circular dichroism  measurements of two coupled monolayers of Chromiun triiodide done in Ref.\cite{Huang2017} reveals that the system is anti-ferromagnetic. Each layer is a ferromagnet with magnetic moments pointing, perpendicular to the layer plane, but in opposite directions. Nonetheless a recent experimental result \cite{Jiang2018}  reveals that an external electric field changes the magnetic state of the bilayer structure, they were able to change the system from an anti-ferromagnetic(AFM) phase to a ferromagnetic(FM) and vice versa.

The electric field control of magnetism has lower energy consumption than current methods in the industry which are based on magnetic fields or electric currents. Understanding the physics behind this kind of control of two dimensional magnetic materials has become a very active line of research \cite{Liu2018,Zhao2018,Lado2017,Matsukura2015,Jiang2018J,Soriano2018} due to the impact it might have in the industry.  

This behavior, known as magneto-electric effect was suggested and measured  \cite{Rontgen,Dzyaloshinski_1959,Fiebig2005} coincidentally in a Chromium anti ferromagnetic crystal ($\rm Cr_{2}O_{3}$). It was observed that an external electric field induces a magnetization \cite{Astrov_1960,Rado1961}, and by the way later in the same system  it was found the inverse, a magnetic field induces a polarization \cite{Astrov_1961}.

\begin{figure}[thpb]
   \centering
{\includegraphics[width=\columnwidth]{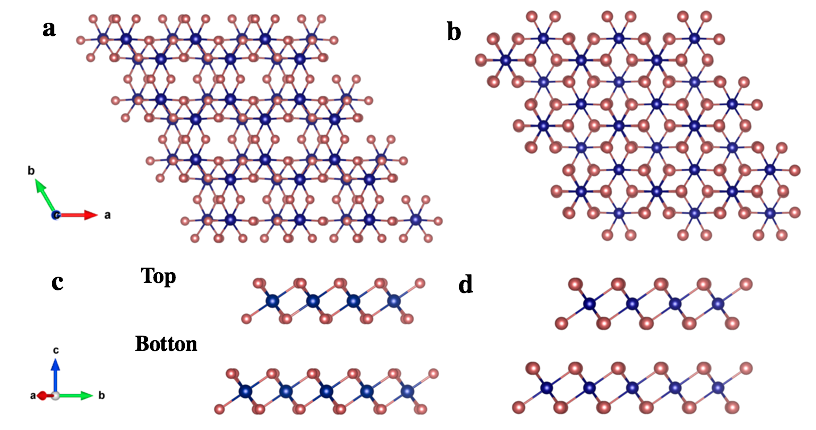}}
\caption{\label{fig:model}
The two possible stacking of $CrI_3$ bilayers. {\bf a}, monoclinic (HT), {\bf b}, Rhomboedral (AB) stacking (LT).
{\bf c,d}, Lateral views. The lattice vectors {\bf a} and {\bf b} indicate the periodic surface unit cell. Cr atoms in blue and I atoms in light red.}
   \end{figure}

In Chromiun triiodide the $\rm Cr^{3+}$ ions are arranged in a honeycomb network surrounded by six $\rm I^{-}$ ions each of them bonded to two Cr atoms. The slabs are stacked with van der Waals forces between them. Two possible stacks have been found in nature, it has been studied that at low temperatures the stacking of $\rm CrI_{3}$ layers is rhombohedral (R3) while for temperatures above 220 K is monoclinic (C2/m) \cite{Mcguire2015}. We will call the rhombohedral stacking, LT and the monoclinic, HT following the notation of Ref. \cite{Jiang2018J}. In Fig.\ref{fig:model} we show the two crystallographic phases.

Recent theoretical calculations shows that the bilayer LT (rhomboedral) phase  remains ferromagnetic while the HT phase becomes an anti-ferromagnet \cite{Soriano2018,Jiang2018J,Sivadas2018}. This dependence on the stacking of the magnetic phase is quite fascinating and reveals that the crystallographic phase measured in Ref.\cite{Jiang2018} is the HT phase albeit it was measured below the transition temperature.



In this article we studied both stacking of Chromiun triiodide monolayers and found with \textit{ab initio} calculations that the rhomboedral (LT) ferromagnetic phase is quite robust and an external electric field cannot modify its magnetic phase, however in the monoclinic (HT) structure the external field induce a phase transition from AFM to FM.

We explain both cases using a simple model of a rigid shift of the energy bands associated with each of the spins. This shift is produced by the external electric field. Our \textit{ab initio} calculations and a toy model reveal that upon the application of the external field the AFM phase gets energy faster than the FM, and for a certain value occurs a magnetic transition in the HT stacking.
In the LT stacking, which was in a FM state  the external field will strengthen the state.
We expect this behavior to be similar in any van der Waals bilayer insulating anti-ferromagnet with opposite magnetization in each layer.


All our calculation were done with Density Functional Theory with Quantum Espresso(QE) \cite{Giannozzi2009}  and corroborated with VASP \cite{vasp1}. In QE the structures were relaxed with projector augmented wave (PAW) pseudopotential \cite{Corso2014} and PBE exchange correlation functional \cite{PBE}, the van der Waals interactions were included for all bilayer structures \cite{Grimme}. On-site Coulomb interaction to the Cr d orbitals was considered with an effective U value of 2.0 eV. A grid of 6$\times$6$\times$1 of kpoints were used to relax the structures and a finer grid of up to 16$\times$16$\times$1 to obtain the total energies, band structures and projected density of states, with a convergence threshold of $10^{-8}$ eV.

\bigskip


We reproduced previous results for monolayer $CrI_{3}$. Our \textit{ab initio}  calculations reveal an energy difference $E_{\rm FM}-E_{\rm AFM}$ of 21 meV/Cr atom. This difference makes the FM state in monolayer $\rm CrI_{3}$ quite robust, and the experimental evidence\cite{Huang2017} confirms that the intralayer ferromagnetism of the bulk system is preserved upon exfoliation. The local magnetic moment of the Cr atoms calculated with QE and VASP is m$\approx$3$\mu_{B}$. The calculation of the magnetic anisotropy energy (MAE) results in an easy axis perpendicular to the plane with a value of 0.7 meV. These results are in agreement with some previous calculations \cite{Lado2017,Zhang2015}. The value of the MAE is close to the ones obtained for different Fe slabs \cite{Li2013}.

Such a high out-of-plane MAE indicates that the magnetic properties of CrI$_3$ monolayers will be preserved in CrI$_3$ bilayer systems; where  we  have two possible interlayer magnetic coupling between the  CrI$_3$ monolayers, {\it i.e.} FM and AFM. Hereafter we will present the total energy results obtained by using the QE code; however the same trend was observed by using the VASP code. We found that,(i) the energy difference between the two crystallographic phases is only 2 meV/Cr atom, being more stable at zero temperature the LT, (ii) the LT stacking exhibits an energetic preference for the  FM state, $E_{\rm AFM}-E_{\rm FM}=3.9$ meV/Cr atom;  whereas (iii) in the HT stacking the AFM phase becomes slightly more favorable,  $E_{\rm AFM}-E_{\rm FM}=-0.17$ meV/Cr atom. It is worth noting that not only the Cr atoms are magnetized, but also the I atoms present net magnetic moment, $\sim$0.2$\rm\mu_B$/I-atom.  For instance, in (i) the I atoms, separated by the vdW interface between the  CrI$_3$ monolayers, present  FM coupling, Cr$^\uparrow$-I$^\downarrow$/I$^\downarrow$-Cr$^\uparrow$; while in (ii) those interface atoms are characterized by an AFM configuration, Cr$^\uparrow$-I$^\downarrow$/I$^\uparrow$Cr$^\downarrow$. Thus, revealing that the interface geometries, of the LT and HT stackings  play an important role on the energetic stability  of the (ground state) FM and AFM phases in CrI$_3$ bilayer systems.

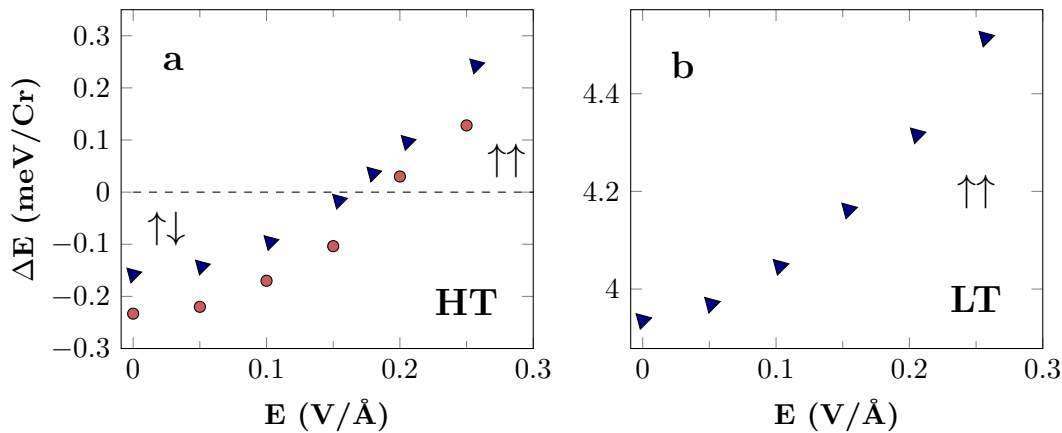
\begin{figure}[thpb]
\input{fig2.tex}
\caption{\label{fig:Field}
Dark blue triangles are \textit{ab initio} calculations of the energy difference between the AFM and FM magnetic phase for {\bf a}, HT and {\bf b}, LT bilayer structures as a function of an external electric field. The light red dots are the results of the rigid bands shift model, explained in the text. 
}
\end{figure}

After corroborating some previous results we concentrate on the effect of the external electric field (EF) on the electronic structure/magnetic phase of the bilayer structures. We found that the EF promotes ferromagnetism in both structures. In the LT stacking, which is FM at zero field, applying an external field makes the difference in energy between FM and AFM phases larger, see Fig.\ref{fig:Field}a, while the HT stacking, which is AFM at zero field, the external EF makes the energy of the FM phase lower than the AFM; this might result in a reversal of the magnetization, it happens for a value around 0.16 \,V/\AA. This result is in agreement with measurements done in Ref.\cite{Jiang2018} although the value we obtain is slightly larger than the one they employed to reverse magnetization in bilayer $\rm CrI_{3}$ (0.08 \,V/\AA). Both values are quite large but encapsulation of the bilayer $\rm CrI_{3}$ in hexagonal boron nitride and graphene as electrodes made possible to apply such a large field. It is relevant to point out that we have checked, relaxing the structure, that an external electric field does not alter significantly the atoms position.

We will try now to explain this behavior and it will be our main result.
To explain the magnetization reversal for the HT phase a first idea suggested in Ref.\cite{Jiang2018} would be that a flow of electrons from one layer to the other, would create an imbalance in the number of electrons per layer. The electrons arriving to a layer, with a different magnetic moment would align their spins due to the strong intralayer exchange coupling and at the end the system will get a net magnetization. We did calculations of the charge transfer and obtain for an EF of 0.25\,V/\AA\, a net charge transfer $\Delta\rho$ = $1.94(1.73)\times 10^{13}$ cm$^{-2}$ between the two layers for FM (AFM) phase , giving rise to a charge density redistribution along the bilayer system; for instance, at the interface region (I atoms) we have  $\Delta\rho$=$\pm 1.06(1.24)\times 10^{13}$cm$^{-2}$, and it increases to   $\pm 3.07(3.04)\times 10^{13}$cm$^{-2}$ at the edge iodine layers; meanwhile  the net charge density of the Cr atoms are slightly modified. The problem with that approach is that the charge is redistributed mainly between the I atoms, there are some changes in I atoms magnetic moment but their magnitudes are small in comparison with Cr magnetic moments, moreover the amount of charge transfer even if it were between Cr atoms would not be able to explain the resulting magnetization in experimental results \cite{Jiang2018}.

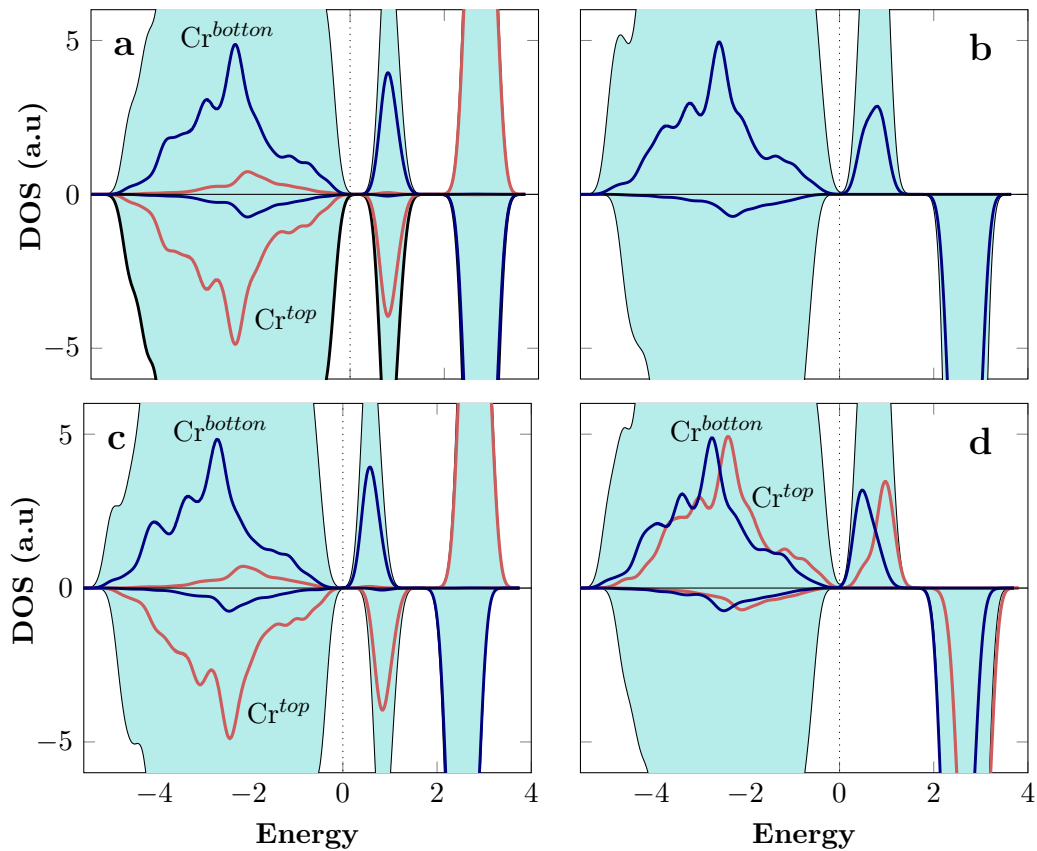
\begin{figure}
  \centering
  \input{fig3.tex}

\caption{\label{fig:pdos}
Total DOS (shaded region), and projected DOS on the Cr-3d orbitals of different layers ($Cr_{top}$ and $Cr_{bottom}$) for the HT structure. In {\bf a, c} the AFM phase with $E=0$ and E=0.25 \,V/\AA\,   respectively. In {\bf b, d} the same for the FM phase.  The Fermi level is set to zero. 
}
   \end{figure}

Lets look now at the total density of states(DOS) and the projected DOS(PDOS) on the Cr atoms for the HT structure. In Fig. \ref{fig:pdos}(a),(b) we show the two phases AFM and FM respectively. We separate the PDOS on Cr atoms on each layer on purpose. When an external EF is applied (Fig. \ref{fig:pdos}(c)-AFM, (d)-FM phase) one notice a shift of the PDOS on Cr atoms, with slight changes in the shape of the curve, the shift is different in each layer or in Cr atoms belonging to different layers. The EF is applied perpendicular to the layers and it will introduce an additional energy to the system, different on each layer. The system will move then their energy levels (bands) in a way to minimize its energy. In the FM phase it will move first the bands associated with the minority spins while in the AFM phase there is no other choice than to move the majority spins of one of the layers. As a result the energy in the AFM systems increases faster than in the FM. To validate our point we calculate the energy of the bands for each phase previous and after turning on the EF from the following expression: 
$E = \int_{-\infty}^{\varepsilon_{\rm F}} \varepsilon \, g(\varepsilon)\, d\varepsilon$ where $g(\varepsilon)$ is the DOS or total PDOS at a given energy $\varepsilon$, $\varepsilon_{\rm F}$ is the Fermi energy, which was obtained integrating the DOS in energy until the correct number of electrons in the system is obtained.  This magnitude plus other terms associated with the energy of the ions, core electrons, etc. should give us the total energy of the system, as we only need differences these other terms should be equal for both phases and they cancel each other. The results are in complete agreement with what we obtained before, the HT system prefers to be AFM at zero field but for large values of the EF there is a phase change. We further exploit this and starting from the zero field DOS (fig.3 (a),(b)) we rigidly shifted by a $\Delta$ one of the bands as explained before, the minority spin bands for the FM and the bands from one of the layer for the AFM phase, we calculate for each configuration the bands energy  and compute the energy difference between phases as a function of $\Delta$. The Fermi level is recalculated for each value of $\Delta$.
The results are the red dots in Fig. \ref{fig:Field} which are in excellent agreement with \textit{ab initio} calculations. This behavior is quite general and we expect it should be valid for any bilayer insulating anti-ferro-magnet when the spins are in opposite directions in each layer.

\begin{figure}[thpb]
\centering
\input{fig4.tex}
\caption{\label{fig:ToyModel}
Model of the DOS for the AFM/FM phase as a function of energy. The zero energy is chosen at the bottom of the bands. An external EF causes the Fermi level to increase and also the total band energy. (a) Both functions, spin up and spin down DOS are equal in the AFM phase, both containing 4 electrons each.  (b) With external EF, an energy shift $\Delta$ between both DOS appears.(c) FM phase  spin up and spin down DOS are different, the D$_{\uparrow}^{\rm FM}$ has now larger number of electrons as compared with the D$_{\downarrow}^{\rm FM}$, such that there is a net magnetic moment of  3$\mu_B$. At zero EF the Fermi level is at $E=1.02$ and both DOS functions have the same width. (d) With an external EF, an energy shift $\Delta$ appears between both DOS. 
}
   \end{figure}
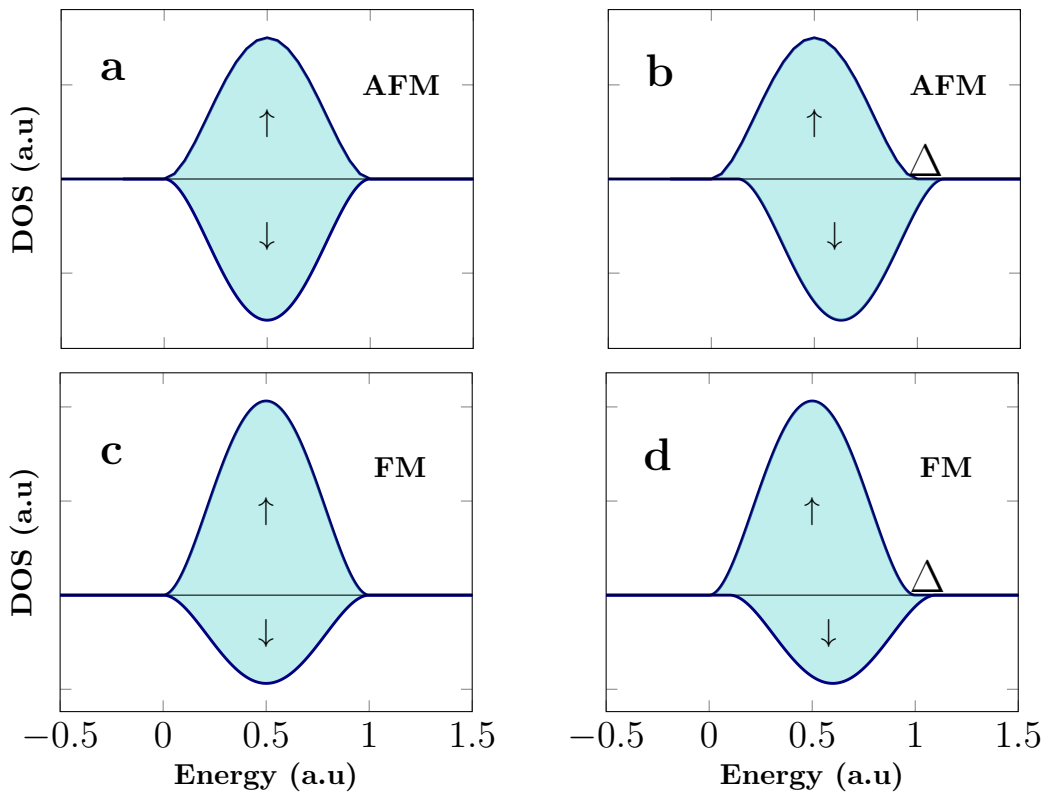

We show now with a toy model how general the above arguments are, it will also clarify more the idea. 
We will simulate the DOS using a simple function with some parameters to have 8 electrons in total. In the AFM phase it would be 4 electrons in each band. The external EF will shift one of the bands by a $\Delta$ as it is shown in Fig. \ref{fig:ToyModel}(a),(b). 

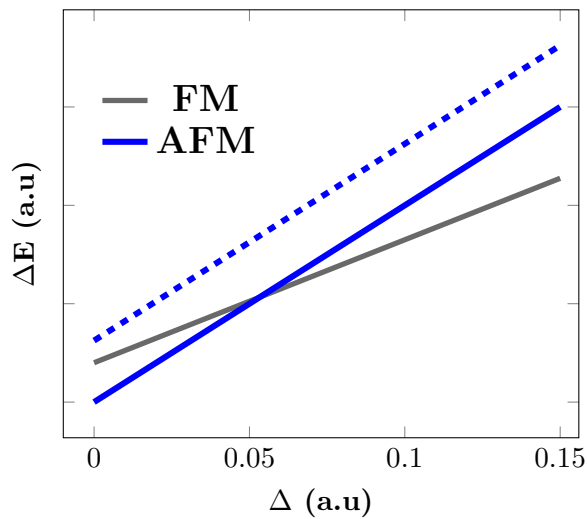
\begin{figure*}[h]
  \centering
  \input{fig_5.tex}
  \caption{Band energy difference (in arbitrary units) between the FM and AFM phase as a function of an energy shift $\Delta$ for the FM and AFM phases.  Black curve depicts the increasing of the energy for the FM phase, and the blue curve for the AFM phase for the HT structure. Above certain critical value of the electric field, (i.e. $\Delta \approx 0.055$)  the FM phase becomes more stable. The dashed blue line(AFM) together with the black line(FM) would correspond to the LT phase.}
  \label{fig:ToyModel2}
\end{figure*}

This rigid shift in energy does not affect the magnetic phase, there are still 4 electrons in each DOS. Therefore no magnetic moment appears, however the shift causes that the total band energy increases, the Fermi level increases also. The total band energy is given by the following expression: 
$E_{\rm AFM}(\Delta) = \int_0^{1+\Delta} \epsilon \left( D_\uparrow^{\rm AFM}(\epsilon)+D_{\downarrow}^{\rm AFM}(\epsilon) \right) d\epsilon$, where $ D_{\uparrow,\downarrow} ^{\rm AFM}(\epsilon)$ are the DOS for spins up and down respectively for a given energy $\epsilon$.
Now we model the FM phase by using the same shape for the DOS functions but with different parameters in such a way to conserve the total number of electrons (8 electrons), but now there are 5.5 and 2.5 electrons in the majority and minority bands respectively. As in the AFM case, the external EF also causes a relative shift, $\Delta$, in energy between the majority and minority DOS. Fig. \ref{fig:ToyModel}(c) shows a pictorial representation for the FM case.
We have a similar expression for the band energy for the FM phase  $E_{\rm FM}(\Delta) = \int_0^{1.02+\Delta} \epsilon \left( D_\uparrow^{\rm FM}(\epsilon)+D_{\downarrow}^{\rm FM}(\epsilon) \right) d\epsilon$ where we have included an extra 0.02 in the upper integration limit.
We did that to mimic a system where at zero EF the AFM phase would be more stable than the FM, to do that we augmented the width of the FM bands by $2\%$, the differences between the band energies calculated for the two phases gives that the AFM is more stable by 0.08 au(arbitrary units). 
In Fig. \ref{fig:ToyModel2} we show the energy difference between the FM phase and the AFM phase as a function of the energy shift $\Delta$. We observe that both, the band energies of the AFM and FM phases increase with $\Delta$, but the energy of the AFM phase increases faster than that of the FM phase.\\
This behavior has the following explanation: in the FM phase the energy shift, caused by the external EF, moves the minority bands toward higher energies, and not the majority ones, because the minority bands contain less electrons (2.5 in this case), therefore the increase in energy is minimized. On the other hand, in the AFM phase, the shift of any DOS (majority or minority) is equivalent because both bands contain the same number of electrons. Therefore the increase in energy is faster for the AFM phase, the slope shown in Fig.\ref{fig:ToyModel2}, there is then a crossing of the curves for a certain value of $\Delta$, thus explaining the AFM to FM transition with electric field.

The same model explains the behavior of the LT crystallographic structure. In Fig.\ref{fig:ToyModel2} the dashed blue line correspond to the AFM state and the black line to the FM state of the LT structure. At $\Delta$=0 the FM phase is more stable and increasing $\Delta$ will only makes the difference in energy between phases larger, it explains why the LT-FM phase of CrI$_3$ monolayer it is so robust under an external EF.

This model does not rely on any particular electronic structure, it will work for any material, in which the minority and majority bands are affected by an EF with a rigid shift of energy bands. The system should be AFM at EF=0 and the difference in energy must be enough for the EF to reverse the phase.
For instance in a bilayer AFM structure made of $\rm VSe_2$ FM monolayers \cite{Gong2018},  the external EF shift the bands associated with one of the spins and the system becomes half metallic. The structure however remains AFM because the energy difference between the two phases is too large.

This shift of the energy bands under an external EF is quite common in van der Waals bilayer materials, for instance in bilayer graphene the system opens a gap and the bands are shifted away from the Fermi level \cite{McCann2007}, in bilayer $\rm MoS_2$ \cite{Liu2012} the bands associated with different spins move separately, in bilayer $\rm InSe$ it is possible to identify the bands associated with different layers by its movement \cite{Shang2018}, etc.

This effect is quite general and one way to express it in a simple case would be a 2$\times$2 Hamiltonian when you add different mass terms in the diagonal, it will open or move the band gap. 
The stacking also plays an important role when two FM monolayers are coupled, the van der Waals interaction is small and also the exchange interlayer coupling. We showed that some changes on the electronic structure on the interface atoms might result in one magnetic phase or the other. 
We expect that more materials displaying magneto-electric effect will appear. The search should be in AFM bulk crystals that might be FM in the monolayer limit. Possible candidates are the recent proposal of monolayer oxyalides $\rm CrOX$ (X=Cl,Br) which are expected to be FM while the bulk materials are AFM \cite{Miao2018} or Transition-Metal Dihydride like $\rm CrH_2$ \cite{Wu2018}.  

In summary we have studied the effect of the electric field on the magnetic phases of bilayer $\rm CrI_{3}$, our \textit{ab initio} calculations reveal that there is a phase transition in the HT structure from AFM to FM for a certain value of the external EF while the LT structure remains in the FM state. We developed a simple model based on a rigid shift of the bands that clearly explains why, under an external EF, a transitions occurs in the HT structure and not in the LT structure. This model can be applied to similar bilayer AFM structures with opposite directions of magnetic moments in each layer. 


\section*{Acknowledgments}

ESM acknowledge financial support from FONDECYT Regular 1170921 (Chile) and thanks Jos\'e Luis Lado for helpful comments. P. Vargas acknowledge support from Financiamiento Basal para Centros Cient\'{\i}ficos y Tecnol\'ogicos de Excelencia, under Project No. FB 0807 (Chile).
RHM  acknowledge financial support from CNPq and FAPEMIG. Powered@NLHPC: This research was partially supported by the supercomputing infrastructure of the NLHPC (ECM-02).
\bigskip



\section*{References}
 \providecommand{\newblock}{}

\bibliographystyle{iopart-num.bst}

\end{document}

%% file: fig2.tex
\definecolor{color1}{RGB}{30,144,255}
\definecolor{gris0}{RGB}{240,240,240}
\definecolor{gris1}{RGB}{230,230,230}
\definecolor{gris2}{RGB}{200,200,200}
\definecolor{gris3}{RGB}{112,128,144}
\definecolor{gris4}{RGB}{120,120,120}
\definecolor{gris5}{RGB}{255,255,240}

\definecolor{hcp}{RGB}{255,255, 0}
\definecolor{fcc}{RGB}{112,138,144}
\definecolor{I4}{RGB}{255,127, 80}
\definecolor{I42}{RGB}{50,205, 50}
\definecolor{cohcp}{RGB}{0,255,0}
\definecolor{ni}{RGB}{255, 20,147}
\definecolor{co1}{RGB}{199, 21,133}
\definecolor{fcc2}{RGB}{0,0,139}
\definecolor{fcc3}{RGB}{255,160,122}
\definecolor{fcc4}{RGB}{72,209,204}
\definecolor{gg}{RGB}{230,230, 230}
\definecolor{indi}{RGB}{75,0,130}
\definecolor{rp}{RGB}{0,255,255}
\definecolor{turq}{RGB}{72,209,204}
\definecolor{indi}{RGB}{75,0,130}
\definecolor{rp}{RGB}{0,255,255}
\definecolor{turq}{RGB}{72,209,204}
\definecolor{coral}{RGB}{255,127,80}
\definecolor{navi}{RGB}{0, 0,128}
\definecolor{green1}{RGB}{51,156,37}
\definecolor{navi}{RGB}{0, 0,128}
\definecolor{viole}{RGB}{199,21,133}
\definecolor{tur1}{RGB}{72,209,204}
\definecolor{sg}{RGB}{173,255, 47}
\definecolor{bon1}{RGB}{250,160,122}
\definecolor{bon2}{RGB}{205, 92, 92}

\begin{tikzpicture}
	\begin{axis}[
	width=0.45\columnwidth,
ymin=-0.3,
ymax=0.35,
xmin=-0.009,
xmax=0.30,
			ytick distance=0.1,
			xtick distance=0.1,
		minor x tick num=1,	xlabel=\small{\textbf{E (V/\AA)}},
ylabel=\small{\textbf{$\Delta$E\hspace{0.001cm} (meV/Cr)}},
			x tick label style={font=\small, /pgf/number format/.cd, set decimal separator={.}, fixed},
			y tick label style={font=\small},
			every axis y label/.style={at={(ticklabel cs:0.5, 5.5)},rotate=90,anchor=center},
legend style={font=,draw=black},
legend entries={},
			legend style={at={(0.25,0.7)},anchor=south,draw=white},
			legend columns=1
		]
\addplot[gray,mark=triangle*, only marks, mark size=3.3pt,mark options={fill=navi,draw=black,rotate=45}] table[x=x,y=y] {
  x        y
0      	-0.157725998724345
0.051422	-0.142527998650621
0.102844	-0.095471998429275
0.154266	-0.016183998923225
0.179977	0.036005999572808
0.205688	0.096525999651931
0.25711	0.243439998303074
0.308532	0.425883998468635

        };           
   \addplot[mark=*,only marks, mark size=2.1pt,mark options={fill=bon2,draw=black,rotate=45}] table[x=x,y=y] {
  x        y
0    -0.233118
0.05 -0.21999999 
0.1   -0.17004729
0.15  -0.10364
0.2    0.0300000000005093
0.25   0.1280000000005093
0.3    1.2799999999997453
        };
        \addplot[dashed,black] coordinates{(0,0) (0.32,0.)};
         \node at (axis cs: 0.28, 0.06) {\large{$\uparrow$$\uparrow$}};
         \node at (axis cs: 0.025,-0.07) {\large{$\uparrow$$\downarrow$}};
		\node at (axis cs: 0.25, -0.21) {\large\textbf{HT}};
        \node at (axis cs: 0.03, 0.25) {\large{\textbf{a}}};
         	\end{axis}
	\end{tikzpicture}
\begin{tikzpicture}
	\begin{axis}[
	width=0.45\columnwidth,
xmin=-0.009,
xmax=0.3,
			xtick distance=0.1,
		minor x tick num=1,	xlabel=\small\textbf{E (V/\AA)},
ylabel={},
			x tick label style={font=\small, /pgf/number format/.cd, set decimal separator={.}, fixed},
			y tick label style={font=\small,},
			every axis y label/.style={at={(ticklabel cs:0.5, 5.5)},rotate=90,anchor=center},
legend style={font=,draw=black},
				legend entries=
{},
			legend style={at={(0.2,0.7)},anchor=south,draw=white},
			legend columns=1
		]
\addplot[mark=triangle*,only marks, mark size=3.5pt,mark options={fill=navi,draw=black,rotate=45}] table[x=x,y=y] {
  x        y
  0               3.9359760007
0.0514220632    3.9693640001
0.1028441264    4.0469180025
0.1542661896    4.1623819998
0.2056882528    4.3163000017
0.257110316     4.5143500007
        }; 
        \node at (axis cs: 0.25, 4.2) {\large{\textbf{$\uparrow$$\uparrow$}}};
		\node at (axis cs: 0.03, 4.46) {\large{\textbf{b}}};
        \node at (axis cs: 0.25, 3.98) {\large\textbf{LT}};
         	\end{axis}
	\end{tikzpicture}

%% file: fig3.tex
\definecolor{color1}{RGB}{30,144,255}
\definecolor{gris0}{RGB}{240,240,240}
\definecolor{gris1}{RGB}{230,230,230}
\definecolor{gris2}{RGB}{200,200,200}
\definecolor{gris3}{RGB}{112,128,144}
\definecolor{gris4}{RGB}{120,120,120}
\definecolor{gris5}{RGB}{255,255,240}
\definecolor{cri}{RGB}{220, 20,60}
\definecolor{lime}{RGB}{46,139,87}

\definecolor{hcp}{RGB}{255,255, 0}
\definecolor{green2}{RGB}{20,235, 34}
\definecolor{green3}{RGB}{34,139, 34}
\definecolor{red2}{RGB}{220,20,60}
\definecolor{coral}{RGB}{255,127, 80}
\definecolor{navi}{RGB}{0, 0,128}
\definecolor{forest}{RGB}{34,139,34}
\definecolor{clari2}{RGB}{176,224,230}
\definecolor{clari}{RGB}{255,222,173}
\definecolor{viole}{RGB}{199,21,133}
\definecolor{green1}{RGB}{51,156,37}
\definecolor{nava}{RGB}{255,255,224}
\definecolor{tur1}{RGB}{72,209,204}
\definecolor{sg}{RGB}{173,255, 47}
\definecolor{bon1}{RGB}{250,160,122}
\definecolor{bon2}{RGB}{205, 92, 92}

\begin{tikzpicture}
		\begin{axis}[
		width=0.48\columnwidth,
		ymin=-6,
		ymax=6,
		xmin=-5.5,
		xmax=4,
xticklabels={},
ylabel=\small{\textbf{DOS (a.u)}},
x tick label style={font=\small},
y tick label style={font=\small},
every axis y label/.style={at={(ticklabel cs:0.5, 5.5)},rotate=90,anchor=center}
		legend style={font=\scriptsize, at={(0.55,0.35)},fill=none ,draw=white},
		every axis y label/.style={at={(ticklabel cs:0.55)},rotate=90,anchor=center}
		]
\addplot[black, fill= tur1!40] table[x=x,y=y1] {1npdos_tot-ht-af-0.dat};
\addplot[black, line width=1.1pt, black, fill= tur1!40] table[x=x,y=y2] {1npdos_tot-ht-af-0.dat};
\addplot[bon2, line width=1.2pt] table[x=x,y=y1] {1ht-af-top-0.dat};
\addplot[bon2, line width=1.2pt] table[x=x,y=y2] {1ht-af-top-0.dat};
\addplot[navi, line width=1.1pt] table[x=x,y=y2] {1ht-af-bot-0.dat};
\addplot[navi, line width=1.1pt] table[x=x,y=y1] {1ht-af-bot-0.dat};
\addplot[black, dotted] coordinates{(0,-8) (0,8)};
\addplot[black] coordinates{(-5,0) (5,0)};
\node at (axis cs: -2.6,5.2){\small{Cr$^{botton}$}};
\node at (axis cs: -1.36,-4.0){\small{Cr$^{top}$}};
\node at (axis cs: -4.8, 4.8){\large{\textbf{a}}};
\end{axis}
\end{tikzpicture}
\begin{tikzpicture}
		\begin{axis}[
		width=0.48\columnwidth,
		ymin=-6,
		ymax=6,
		xmin=-5.5,
		xmax=4,
ylabel={},
xticklabels={},
yticklabels={},
x tick label style={font=\scriptsize},
y tick label style={font=\scriptsize},
every axis y label/.style={at={(ticklabel cs:0.5, 5.5)},rotate=90,anchor=center}
		x tick label style={font=\scriptsize},
		y tick label style={font=\scriptsize},
		legend style={font=\scriptsize, at={(0.55,0.35)},fill=none ,draw=white},
		every axis y label/.style={at={(ticklabel cs:0.55)},rotate=90,anchor=center}
		]
\addplot[black, fill= tur1!40] table[x=x,y=y1] {1npdos_tot-ht-fm-0.dat };
\addplot[black, fill= tur1!40] table[x=x,y=y2] {1npdos_tot-ht-fm-0.dat };
\addplot[navi, line width=1.1pt] table[x=x,y=y2] {1ht-fm-0-bot.dat};
\addplot[navi, line width=1.1pt] table[x=x,y=y1] {1ht-fm-0-bot.dat};
\addplot[black, dotted] coordinates{(0,-8) (0,8)};
\addplot[black] coordinates{(-5,0) (5,0)};
\node at (axis cs: 3.0, 4.8){\textbf{\large{b}}};
\end{axis}
\end{tikzpicture}
\begin{tikzpicture}
		\begin{axis}[
		width=0.48\columnwidth,
		ymin=-6,
		ymax=6,
		xmin=-5.5,
		xmax=4,
xlabel=\textbf{\small{Energy}},
ylabel=\small{\textbf{DOS (a.u)}},
x tick label style={font=\small},
y tick label style={font=\small},
every axis y label/.style={at={(ticklabel cs:0.5, 5.5)},rotate=90,anchor=center}
		legend style={font=\scriptsize, at={(0.55,0.35)},fill=none ,draw=white},
		every axis y label/.style={at={(ticklabel cs:0.55)},rotate=90,anchor=center}
		]
\addplot[black,  fill= tur1!40] table[x=x,y=y1] {2pdos_tot-ht-af-e.dat};
\addplot[black,  fill= tur1!40] table[x=x,y=y2] {2pdos_tot-ht-af-e.dat};
\addplot[bon2, line width=1.2pt] table[x=x,y=y2] {2ht-af-top-e.dat};
\addplot[bon2, line width=1.2pt] table[x=x,y=y1] {2ht-af-top-e.dat};
\addplot[navi, line width=1.1pt] table[x=x,y=y2] {2ht-af-bot-e.dat};
\addplot[navi, line width=1.1pt] table[x=x,y=y1] {2ht-af-bot-e.dat};
\addplot[black,dotted] coordinates{(0,-8) (0,8)};
\addplot[black] coordinates{(-5,0) (5,0)};
\node at (axis cs: -2.6,5.2){\small{Cr$^{botton}$}};
\node at (axis cs: -1.36,-4.0){\small{Cr$^{top}$}};
\node at (axis cs: -4.8, 4.8){\textbf{\large{c}}};;
\end{axis}
\end{tikzpicture}
\begin{tikzpicture}
		\begin{axis}[
		width=0.48\columnwidth,
		ymin=-6,
		ymax=6,
		xmin=-5.5,
		xmax=4,
xlabel=\textbf{\small{Energy}},
yticklabels={}
ylabel={},
x tick label style={font=\small},
y tick label style={font=\small},
every axis y label/.style={at={(ticklabel cs:0.5, 5.5)},rotate=90,anchor=center}
		legend style={font=\scriptsize, at={(0.55,0.35)},fill=none ,draw=white},
		every axis y label/.style={at={(ticklabel cs:0.55)},rotate=90,anchor=center}
		]
\addplot[black,  fill= tur1!40] table[x=x,y=y1] {3pdos_tot-ht-fm-e.dat};
\addplot[black,  fill= tur1!40] table[x=x,y=y2] {3pdos_tot-ht-fm-e.dat};
\addplot[bon2,  line width=1.2pt] table[x=x,y=y2] {3ht-fm-top-e.dat};
\addplot[bon2,  line width=1.2pt] table[x=x,y=y1] {3ht-fm-top-e.dat};
\addplot[navi, line width=1.1pt] table[x=x,y=y2] {3ht-fm-bot-e.dat};
\addplot[navi, line width=1.1pt] table[x=x,y=y1] {3ht-fm-bot-e.dat};
\addplot[black,dotted] coordinates{(0,-8) (0,8)};
\addplot[black] coordinates{(-5,0) (5,0)};
\node at (axis cs: -2.6,5.2){\small{Cr$^{botton}$}};
\node at (axis cs: -1.2,3.0){\small{Cr$^{top}$}};
\node at (axis cs: 3.0, 4.8){\textbf{\large{d}}};
\end{axis}
\end{tikzpicture}

%% file: fig4.tex
\definecolor{san}{RGB}{255,160,122}
\definecolor{blu}{RGB}{173,216,230}
\definecolor{coral}{RGB}{255,127, 80}
\definecolor{indi}{RGB}{75,0,130}
\definecolor{medtur}{RGB}{72,209,204}
\definecolor{gris69}{RGB}{105,105,105}
\definecolor{gris69}{RGB}{105,105,105}
\definecolor{green1}{RGB}{51,156,37}
\definecolor{yel}{RGB}{250,235,215}
\definecolor{mag}{RGB}{199,21,85}
\definecolor{bon}{RGB}{30,144,255}

\definecolor{navi}{RGB}{0, 0,128}
\definecolor{viole}{RGB}{199,21,133}
\definecolor{tur1}{RGB}{72,209,204}
\definecolor{sg}{RGB}{173,255, 47}

\hspace{-0.55cm}
\begin{tikzpicture}
\begin{axis}[
width=0.45\columnwidth,
xmin=-0.5,
xmax=1.5,
xlabel={},
xticklabels={},
yticklabels={},
ylabel=\textbf{\small{DOS (a.u)}},
y tick label style={font=\large},
every axis y label/.style={at={(ticklabel cs:0.5, 5.5)},rotate=90,anchor=center}
]
\addplot[navi,line width=1.1pt] table[x=x,y=y]{up1.txt};
\addplot[fill=tur1,opacity=0.35] table[x=x,y=y]{up1.txt};
\addplot[fill= tur1,opacity=0.35]
 table[x=x,y=y]{dw.dat};
 \addplot[navi,line width=1.1pt] table[x=x,y=y]{dw.dat};
 \addplot[black] coordinates{(-5,0) (5,0)};
\node at (axis cs: 0.5, 3){$\big\uparrow$};
\node at (axis cs: 0.5, -3){$\big\downarrow$};
\node at (axis cs:1.15,5){\small{\textbf{AFM}}};
\node at (axis cs:-0.25,5.8){\Large{\textbf{a}}};
\end{axis}
\end{tikzpicture}
\hspace{1.1cm}
\begin{tikzpicture}
\begin{axis}[
width=0.45\columnwidth,
xmin=-0.5,
xmax=1.5,
xlabel=\textbf{Energy},
xlabel={},
xticklabels={},
ylabel={},
yticklabels={}
every axis y label/.style={at={(ticklabel cs:0.5, 5.5)},rotate=90,anchor=center}
]
\addplot[navi,line width=1.1pt] table[x=x,y=y]{up1.txt};
\addplot[fill=tur1,opacity=0.35] table[x=x,y=y]{up1.txt};
\addplot[navi,line width=1.1pt] table[x=x,y=y]{dw4.dat};
\addplot[fill= tur1,opacity=0.35] table[x=x,y=y]{dw4.dat};
\addplot[black] coordinates{(-5,0) (5,0)};
\node at (axis cs: 0.5, 3){$\big\uparrow$};
\node at (axis cs: 0.6, -3){$\big\downarrow$};
\node at (axis cs:1.15,5){\small{\textbf{AFM}}};
\node at (axis cs: 1.04, 1){\Large{\textbf{$\Delta$}}};
\node at (axis cs:-0.25,5.8){\Large{\textbf{b}}};
\end{axis}
\end{tikzpicture}
\hspace{0.3cm}
\begin{tikzpicture}
\begin{axis}[
width=0.45\columnwidth,
xmin=-0.5,
xmax=1.5,
xlabel=\textbf{\small {Energy (a.u)}},
ylabel=\textbf{\small{DOS (a.u)}},
yticklabels={},
x tick label style={font=\large},
y tick label style={font=\large},
every axis y label/.style={at={(ticklabel cs:0.5, 5.5)},rotate=90,anchor=center}
]
\addplot[navi,line width=1.1pt] table[x=x,y=y]{up3.dat};
\addplot[fill= tur1,opacity=0.35] table[x=x,y=y]{up3.dat};
\addplot[fill= tur1,opacity=0.35] table[x=x,y=y]{dw3.dat};
\addplot[navi,line width=1.1pt] table[x=x,y=y]{dw3.dat};
\addplot[black] coordinates{(-5,0) (5,0)};
\node at (axis cs: 0.5, 4.5){$\big\uparrow$};
\node at (axis cs: 0.5, -2){$\big\downarrow$};
\node at (axis cs:1.15,6.8){\small{\textbf{FM}}};
\node at (axis cs:-0.25,7.5){\Large{\textbf{c}}};
\end{axis}
\end{tikzpicture}
\hspace{0.4cm}
\begin{tikzpicture}
\begin{axis}[
width=0.45\columnwidth,
xmin=-0.5,
xmax=1.5,
xlabel=\textbf{\small{Energy (a.u)}},
yticklabels={}
ylabel={},
x tick label style={font=\large},
y tick label style={font=\large},
every axis y label/.style={at={(ticklabel cs:0.5, 5.5)},rotate=90,anchor=center}
]
\addplot[navi,line width=1.1pt] table[x=x,y=y]{up3.dat};
\addplot[fill=tur1,opacity=0.35] table[x=x,y=y]{up3.dat};
\addplot[fill=tur1,opacity=0.35] table[x=x,y=y]{dw2.dat};
\addplot[navi,line width=1.1pt] table[x=x,y=y]{dw2.dat};
\addplot[black] coordinates{(-5,0) (5,0)};
\node at (axis cs: 0.5, 4.5){$\big\uparrow$};
\node at (axis cs: 0.58, -2){$\big\downarrow$};
\node at (axis cs: 1.06, 1){\Large{\textbf{$\Delta$}}};
\node at (axis cs:1.15,6.8){\small{\textbf{FM}}};
\node at (axis cs:-0.25,7.5){\Large{\textbf{d}}};
\end{axis}
\end{tikzpicture}

%% file: fig_5.tex
\definecolor{san}{RGB}{255,160,122}
\definecolor{blu}{RGB}{173,216,230}
\definecolor{coral}{RGB}{255,127, 80}
\definecolor{indi}{RGB}{75,0,130}
\definecolor{medtur}{RGB}{72,209,204}
\definecolor{gris69}{RGB}{105,105,105}
\definecolor{gris69}{RGB}{105,105,105}
\definecolor{green1}{RGB}{51,156,37}
\definecolor{yel}{RGB}{250,235,215}
\definecolor{mag}{RGB}{199,21,85}
\definecolor{bon}{RGB}{30,144,255}
\definecolor{navi}{RGB}{0, 0,128}
\definecolor{viole}{RGB}{199,21,133}
\definecolor{tur1}{RGB}{72,209,204}
\definecolor{sg}{RGB}{173,255, 47}

\begin{tikzpicture}
\begin{axis}[
xmin=-0.0099,
xmax=0.156,
xtick distance=0.05,
xlabel=\textbf{\small{$\Delta$ (a.u)}},
yticklabels={},
ylabel=\textbf{\small{$\Delta$E (a.u)}},
x tick label style={font=\small,/pgf/number format/.cd, set decimal separator={.}, fixed},
y tick label style={font=\small},
every axis y label/.style={at={(ticklabel cs:0.5, 5.9)},rotate=90,anchor=center},
legend style={at={(0.4,0.85)},draw=white}
]
\addplot[domain=0:0.15,gris69,line width=2.0pt]
{4.08 + 2.5*x};
\addplot[domain=0:0.15,blue,line width=2.3pt]
{4.0 + 4*x};
\addplot[domain=0:0.15,blue,line width=2.3pt,dashed]
{4.125 + 4*x};
\legend{\textbf{\large{FM}}, \textbf{\large{AFM}}}
\end{axis}
\end{tikzpicture}